\documentclass[journal=jacsat,manuscript=article]{achemso}
\usepackage[version=3]{mhchem} 

\usepackage{graphicx}
\usepackage{dcolumn}
\usepackage{bm, bbm}
\usepackage{xcolor}
\usepackage{float}
\usepackage{comment}
\usepackage{mdwtab}

\newcommand{\ket}{{\rangle}}
\newcommand{\bra}{{\langle}}

\newcommand{\blue}{}

\title{Mixed Planewave and Localized Orbital Basis for Sparse-Stochastic Hybrid TDDFT}

\author{Kyle Chen}
\altaffiliation{These authors contributed equally.}
\author{Barry Y. Li}
\altaffiliation{These authors contributed equally.}
\author{Tucker Allen}
\author{Daniel Neuhauser} 
\email{dxn@ucla.edu}
\affiliation[University of California, Los Angeles]
{Department of Chemistry and Biochemistry, University of California, Los Angeles, Los Angeles, California 90095, United States}

\begin{document}

\begin{tocentry}
\begin{figure}[H]
\centering
\includegraphics[width=3in]{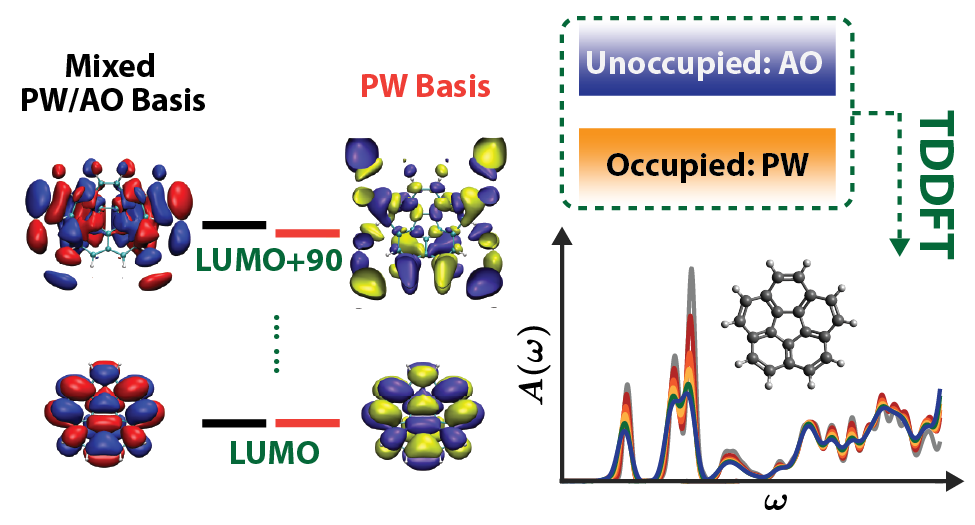}
\end{figure}
\end{tocentry}

\begin{abstract}
We present a mixed basis-set approach to obtain optical absorption spectra within a generalized Kohn-Sham time-dependent density functional theory framework. All occupied valence molecular orbitals (MOs) are expanded in a plane-wave (PW) basis, while unoccupied MOs are derived primarily from localized atomic basis functions. The method accelerates spectral convergence when compared to fully PW-based simulations, with a $2-3$ fold reduction in the number of unoccupied MOs entering the Casida equation. The mixed-basis is placed on a common real-space grid, enabling our previously developed deterministic/sparse-stochastic evaluation of the exact exchange operator (\blue{\textit{J. Chem. Theory Comput.} \textbf{2023}, \textit{19}, 9239–9247}). This chemically intuitive and computationally efficient approach is validated across various molecular systems, including $\pi$-conjugated polymethine dyes, aromatic hydrocarbons, and a chlorophyll monomer.
\end{abstract}

\section{Introduction}
Time-dependent density functional theory (TDDFT) is a widely used method for the excited-state properties of molecules, known for possessing a remarkable balance between accuracy and computational efficiency. An extension of ground-state DFT, TDDFT is built upon the Runge-Gross theorem which reduces the time evolution of the many-body electronic wavefunction to the independent propagation of single-particle molecular orbitals (MOs).~\cite{RungeGross1984} Since its inception, TDDFT has been used in predicting both linear and nonlinear optical response, photoelectron spectra, and optoelectronic properties in general.\cite{casida1995,jamorski1996,yabana1997,yabana1999,lopata2012,debeer2008,lerme1998,qian2006,baer2003,baer2001,baer2004,zelovich2014,ding2013} Modern implementations of TDDFT employ hybrid functionals that incorporate a fraction of Fock exchange to the exchange-correlation (XC) potential. Inclusion of exact exchange provides a better description of delocalized excited states, such as $\pi\rightarrow\pi^*$ transitions, charge-transfer excitations, and excitations in systems with pronounced excitonic effects.\cite{becke1993,baer2010,leininger1997,Yanai2004,karolewski2011,kronik2012,vlcek2019} 

Going beyond the local density and generalized gradient approximations (LDA/GGA)~\cite{PerdewWang1992,perdew_generalized_1996} offers improved accuracy but adds to the computational cost with the need to evaluate the exchange integrals. In previous work, we introduced a mixed deterministic/sparse-stochastic exchange approach that reduces the scaling of both hybrid DFT and linear-response TDDFT within a plane-wave (PW) representation.\cite{bradbury_neargap_2023,Sereda2024} This method splits the exchange interaction kernel, $u(k)$, into low- and high-$k$ components (where $k$ denotes a reciprocal lattice vector): 
\begin{equation}
\begin{aligned}
    u(k) &= \sum_{k_{\text{low}}} |k_{\text{low}}\rangle u(k_{\text{low}}) \langle k_{\text{low}}| \\
    &+ \sum_{k_{\text{high}}} \sqrt{u(k_{\text{high}})} |k_{\text{high}}\rangle \langle k_{\text{high}}| \sqrt{u(k_{\text{high}})}.
\end{aligned}
\end{equation}
The low-$k$ components are treated exactly while the high-$k$ terms are represented by short (sparse) random vectors that uniformly sample the high-$k$ space. The cutoff momentum, $k_{\text{cut}}$, that separates the low- and high-$k$ spaces is a convergence parameter. The size of the stochastic basis sampling the high-$k$ space is system-size independent, enabling large-scale hybrid TDDFT of molecular complexes with thousands of valence electrons. 

In previous sparse-stochastic hybrid TDDFT calculations ~\cite{Sereda2024}, we found that many unoccupied MOs (roughly three to four times the number of occupied states) are required to obtain a converged absorption spectrum. This issue of convergence is well-known and inherent to the use of a PW basis-set.~\cite{Booth2016,Sun2017} In this work, we resolve this issue by implementing a mixed plane-wave/atomic-orbital (PW/AO) basis-set representation for the MOs. The integration of mixed basis sets, combining PWs and AOs, is very useful in computational chemistry and materials science.\cite{Colle1987,Euwema1971,Fabian2025,Carsky1998,liu2015,ko2021} For example, Samsonidze et al. demonstrated that unoccupied orbitals in GW calculations could be replaced with simple approximate physical orbitals, where Gaussian orbitals are used for resonant and continuum states.\cite{Samsonidze2011} Similarly, works conducted by Booth et al. and Sun et al. focus on Gaussian and PW hybrid representations alongside density fitting techniques.\cite{Booth2016,Sun2017}

We now turn to the primary focus of this article: how to reduce the number of unoccupied states entering the Casida equation through a mixed basis-set representation. Below, we present a PW/AO mixed-basis approach and its TDDFT applications on $\pi$-conjugated flavylium (Flav-9) and indocyanine green (ICG-7) dyes, curved aromatic corannulene (C$_{20}$H$_{10}$), as well as a monomer chlorophyll a (Chla) complex.

\section{Methodology}

Here we are interested in the optical absorption spectra of molecules using a mixed PW/AO basis-set representation for the MOs. In Section A, an iterative Chebyshev approach for efficiently calculating spectra is presented. In Section B, the mixing of PW and AO bases is outlined, and in Section C we show an orthonormalization procedure for the combined basis-set MOs. Finally, in Section D we provide metrics for convergence of the TDDFT spectra.

\subsection{A. Iterative approach for optical absorption spectra}
Excitation energies are obtained by solving the Casida equation~\cite{casida1995}:
\begin{equation}
    \mathcal{L}\begin{pmatrix}f^+ \\ f^-\end{pmatrix} = \omega\begin{pmatrix}1 & 0 \\ 0 & -1\end{pmatrix}\begin{pmatrix}f^+ \\ f^-\end{pmatrix},
\end{equation}
where the Liouvillian is
\begin{equation}
\label{casida}
    \mathcal{L}=\blue{\begin{pmatrix}\mathcal{A} & \mathcal{B} \\ -\mathcal{B} & -\mathcal{A}\end{pmatrix}},
\end{equation}
with $f^+$ and $f^-$ transition eigenvectors corresponding to excitations and de-excitations, respectively. For single-particle spin-singlet excitations, the $\blue{\mathcal{A}}$ and $\blue{\mathcal{B}}$ matrices read (assuming real-valued MOs throughout):
\begin{equation}
    \begin{aligned}
        \blue{\mathcal{A}_{ia,jb}} &= (\varepsilon_a - \varepsilon_i)\delta_{ij}\delta_{ab} + 2(ia|jb)\\ &+ (ia|f_{\text{XC}}|jb)- (\phi_a\phi_b|u|\phi_i\phi_j),
    \end{aligned}
\end{equation}
\begin{equation}
    \blue{\mathcal{B}_{ia,bj}} = 2(ia|bj) + (ia|f_{\text{XC}}|bj)- (\phi_a\phi_j|u|\phi_i\phi_b),
\end{equation}
where $\phi_{i,j,...}$ denote occupied (valence) generalized Kohn-Sham (GKS) MOs with associated energies $\varepsilon_{i,j,...}$ and $\phi_{a,b,...}$ unoccupied (conduction) GKS MOs with energies $\varepsilon_{a,b,...}$.

The Hartree integrals are: 
\begin{equation}
    (ia|jb) = \int dr \, dr' \, \phi_i(r) \phi_a(r) |r-r'|^{-1} \phi_j(r') \phi_b(r').
\end{equation}
The matrix elements of the XC kernel, $f_{\text{XC}}$, are evaluated to first-order within an adiabatic LDA scheme.~\cite{Baroni1987} The exchange integrals are calculated under a kernel $u$ that is split to short- and long-range parts in real space \cite{becke1993,savin1997}:
\begin{equation}
\label{rsr}
    \begin{aligned}
        u(|r-r'|) &= \frac{1-(\alpha+\beta \cdot  \text{erf} (\gamma|r-r'|))}{|r-r'|} \\ &+\frac{\alpha+\beta \cdot \text{erf} (\gamma|r-r'|)}{|r-r'|}.
    \end{aligned}
\end{equation}

Rather than direct diagonalization of Eq. (\ref{casida}), we use an iterative Chebyshev polynomial expansion approach to access the eigenvalues of $\mathcal{L}$, as in Refs.~\cite{bradbury_bethesalpeter_2022,bradbury_optimized_2023,Sereda2024,baer2004}. The absorption spectrum is calculated as: 
\begin{equation}
A(\omega)\propto\omega\bra \chi|\delta(\mathcal{L}-\omega)|\chi\ket,
\end{equation}
where 
\begin{equation}
    |\chi\ket = \begin{pmatrix}
    \chi_{ia}^+ \\
    \chi_{ia}^-
    \end{pmatrix} = \begin{pmatrix}
    +\langle\phi_a|\hat{r}|\phi_i\rangle \\
    -\langle\phi_a|\hat{r}|\phi_i\rangle
    \end{pmatrix},
\end{equation}
is a single exciton vector corresponding to a field polarization in the $\hat{r}$ direction, \blue{ combining forward ($+$) and backward ($-$) dipole transitions as detailed in Refs.\cite{baer2004,Sereda2024}} The Gaussian-broadened delta function is expanded in terms of Chebyshev polynomials: 
\begin{equation}
\delta(\mathcal{L} - \omega) = \sum_{n=0}^{N_{\text{Chebyshev}}} c_n(\omega) T_n(\tilde{\mathcal{L}}),
\end{equation}
where $T_n$ is the $n$'th-order Chebyshev polynomial and $\tilde{\mathcal{L}}$ is a
scaled Liouvillian with eigenvalues between $-1$ and $1$.\cite{bradbury_optimized_2023,Weisse2006} $A(\omega)$ is therefore calculated from the Chebyshev moments,
\begin{equation}
A(\omega) = \frac{4\pi\omega}{c}\sum_{n=0}^{N_{\text{Chebyshev}}} c_n(\omega)\bra\tilde{\chi}|T_n(\tilde{\mathcal{L}})|\chi\ket,
\end{equation}
where $|\tilde{\chi}\ket =\pm|\chi\ket$ and for the coefficients, $c_n(\omega)$, simple smoothly decaying weights are used.\cite{Weisse2006}

\subsection{B. Mixed-basis MOs on a real-space grid}
First, we perform two separate LDA-DFT calculations, one using a PW basis and the other with a localized AO basis. All vacuum molecular geometries are optimized in the ORCA 6.0 program at the B3LYP/def2-TZVPPD level of theory.\cite{Neese2020,Neese2022} The PW calculations are performed using an in-house DFT code with norm-conserving pseudopotentials (NCPPs) representing the core electrons.~\cite{TroullierMartins91} The all-electron AO-basis calculations utilize the $PySCF$ package.\cite{Sun2018} The choice of basis is guided by benchmark calculations on the C$_{20}$H$_{10}$ molecule, where we compare the lowest-occupied MO (LUMO) - highest-occupied MO (HOMO) energy gaps obtained from LDA-DFT using various basis sets against a reference grid-based PW-LDA-DFT calculation. Among the tested basis-sets, aug-cc-pVDZ yields the smallest deviation from the PW result, as detailed in Appendix A.\cite{dunning2001}

\begin{figure}[H]
\centering
\includegraphics[width=3in]{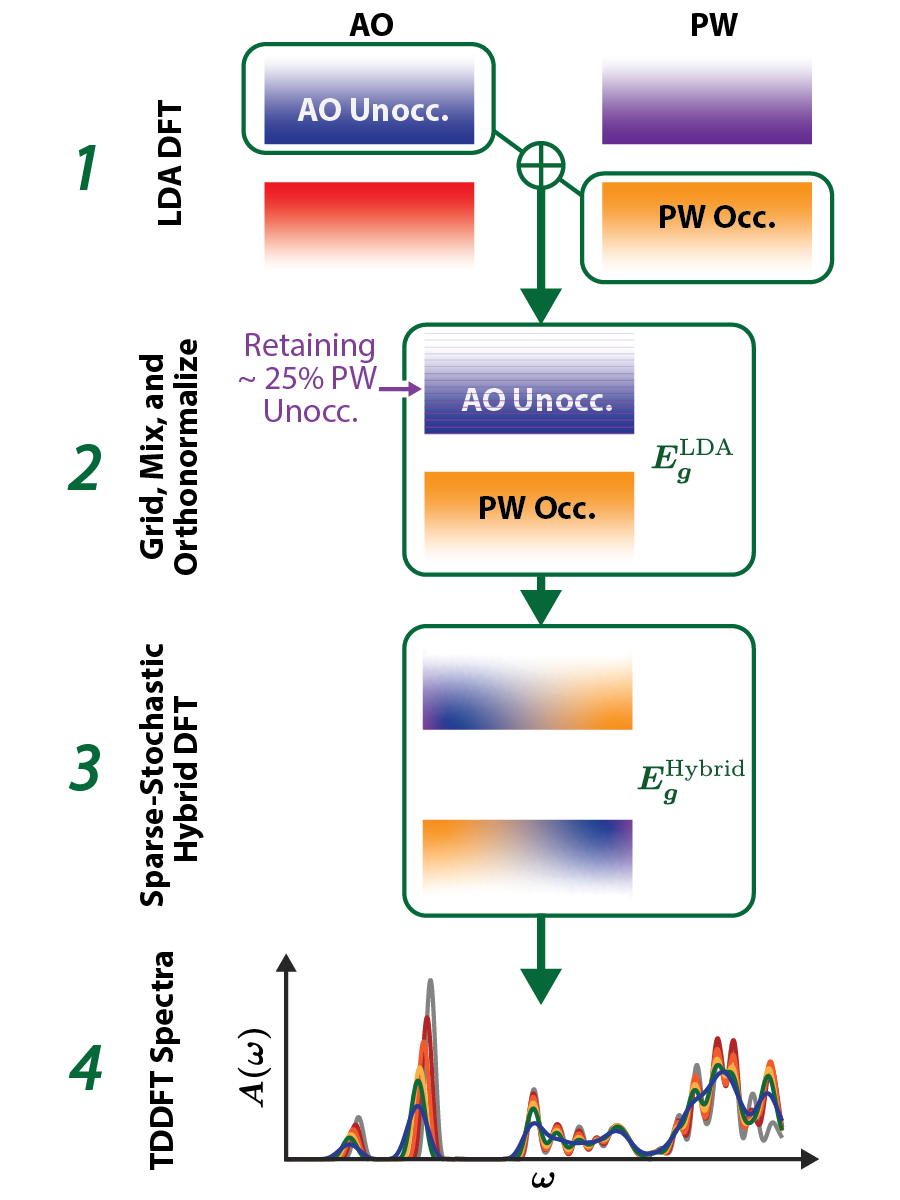}
\caption{\label{fig:fig_1} Schematic representation of this work.}
\end{figure}

In the PW-LDA-DFT calculations, the molecule is centered on a uniform real-space grid with a spacing of $dx=dy=dz=0.4$ Bohr. The simulation grids include at least 6 Bohr of padding beyond the extent of the molecule in all directions, \blue{this padding ensures proper boundary conditions and avoids periodic images in finite systems.} The Martyna-Tuckerman approach is used to further minimize edge effects.~\cite{MartynaTuckerman1999}. Both the PW- and AO-MOs are then evaluated on this same grid. By replacing all or a subset of the unoccupied PW-MOs with those obtained from the AO calculation, we construct a mixed-basis representation of the MOs (Fig. \ref{fig:fig_1}). 

For static bandgaps we find that all PW-based unoccupied states can be replaced with a lesser number of AO-derived MOs and still reproduce the gaps of the fully-PW based calculations. However, a more careful selection of the virtual-space MOs is required for excited-state simulations. For sparse-stochastic hybrid TDDFT calculations, we observe that retaining every fourth unoccupied MO from the PW basis while replacing the rest with AO-derived orbitals provides the best agreement with the first excitation energies, i.e., absorption onset, obtained from fully PW-TDDFT.  Specifically, the AO basis sets tend to overly localize, particularly in the virtual space, often resulting in a basis-set-induced blueshift. By retaining approximately 25\% of the PW unoccupied MOs, we mitigate this shift and recover excitation energies consistent with fully PW-based calculations. \blue{The 25\% PW retaining ratio is optimized for molecular systems exhibiting naturally localized transitions, where our method demonstrates particular strength.} To ensure a consistent set of MOs, we apply this same mixing strategy in both the static DFT and excited-state TDDFT simulations.

Further, the mixing of an all-electron AO basis-set with an NCPP-PW approach results in lower HOMO energies in the PW-LDA-DFT calculations. To ensure energy alignment between the two frameworks, we apply a scissor-shift correction ($\Delta\varepsilon$), i.e., $\varepsilon_{i}^{PW} = \varepsilon_{i}^{AO} + \Delta \varepsilon$, where $\Delta\varepsilon$ is determined from the difference between the HOMO energies. \cite{Samsonidze2011} \blue{The HOMO energy alignment accounts for systematic differences between the all-electron AO and NCPP-PW treatments. Although the absolute energies differ due to their distinct representations of the electron-nuclear interaction, the preservation of energy gaps ensures the physical validity of the mixed-basis approach.} \blue{The initial PW/AO orbital mixing occurs at the LDA level without self-consistency, requiring only HOMO alignment. We then use the mixed basis to perform the GKS-DFT SCF procedure.}

\subsection{C. Orthonormalization} 
Subsequently, we perform an overall orthonormalization of the mixed-basis MOs using QR decomposition by Householder transformation.\cite{golub1965,press2007} This is done by decomposing the eigenvector matrix $\sqrt{dV}C$, where the columns of $C$ are non-orthogonal mixed LDA-DFT MOs and $dV$ is the volume element, into an orthonormal matrix $Q$ and an upper triangular matrix $R$. The columns of $Q$ correspond to the orthonormalized MOs on the grid. While other orthonormalization techniques, such as Gram-Schmidt or Löwdin transformation with singular value decomposition (SVD),\cite{szabo1996,daniel1976,golub1970,lowdin1950,bowler2012} are also applicable, we find the QR method via Householder transformation to be the most efficient. A Frobenius norm error analysis comparing different orthonormalization techniques is shown in Appendix B. 

\subsection{D. TDDFT Spectra and Spectral Convergence}
With orthonormalized mixed-basis LDA MOs \blue{placed on a common real-space grid}, we proceed with a sparse-stochastic hybrid DFT calculation using the CAM-LDA0 functional ($\alpha=0.19,\beta=0.46,\gamma=0.33$ $\rm{Bohr}^{-1}$).\cite{Yanai2004,bradbury_neargap_2023,Sereda2024} This calculation updates the LDA-DFT Hamiltonian to include parametrized exact-exchange within a GKS-DFT framework. The \blue{high-$k$ components of the} hybrid exchange operator are represented with 1000 short stochastic fragments. \blue{The required number of stochastic fragments to obtain converged bandgaps and spectra (and the associated small stochastic error) has been extensively characterized in our previous work;} for further details, see Ref.~\cite{bradbury_neargap_2023, Sereda2024}. \blue{As the GKS-DFT calculation updates mixed-basis LDA MOs self-consistently, 1000 stochastic fragments uniformly sample both basis types while preserving their characters through the real-space representation.} Subsequently, we perform sparse-stochastic hybrid TDDFT calculations with 1000 Chebyshev terms to compute the optical absorption spectra. All occupied valence electron MOs are included in the valence space ($N_v = N_{\text{occ}}$), while in the TDDFT calculations the number of unoccupied MOs ($N_c$) is varied. \blue{All TDDFT simulations use  converged GKS-DFT MOs and eigenvalues within the mixed PW/AO representation as inputs into the Casida equation.}

To quantify the convergence of optical spectra with respect to $N_c$, we define the deviation matrix $D$,
\begin{equation}
    D_{ij}=\int|\tilde{A}_i(\omega)-\tilde{A}_j(\omega)|\;d\omega,
\end{equation}
where $\tilde{A}(\omega)$ is the area-normalized absorption spectrum, $A(\omega)$, i.e., $\tilde{A}(\omega)=A(\omega)/\int A(\omega)\,d\omega$. Consequently, $D_{ij}$ ranges between 0 (perfectly identical spectra) and 2 (completely different spectra). We further define two averaged quantities $\delta$ and $\bar{\delta}$,  
\begin{align}
    \delta &= \frac{1}{N-1} \sum_{i=1}^{N-1} D_{i,i+1}, \\
    \bar{\delta} &= \frac{2}{N(N-1)} \sum_{i<j} D_{ij},
\end{align}
where $N$ denotes the total number of spectra, each corresponding to a different value of $N_c$ included in the TDDFT calculation. The quantity $\delta$ measures the average deviation between successive spectra, while $\bar{\delta}$ captures the average deviation over all pairs.

\section{Results and Discussion}
\begin{figure}[H]
\centering
\includegraphics[width=3in]{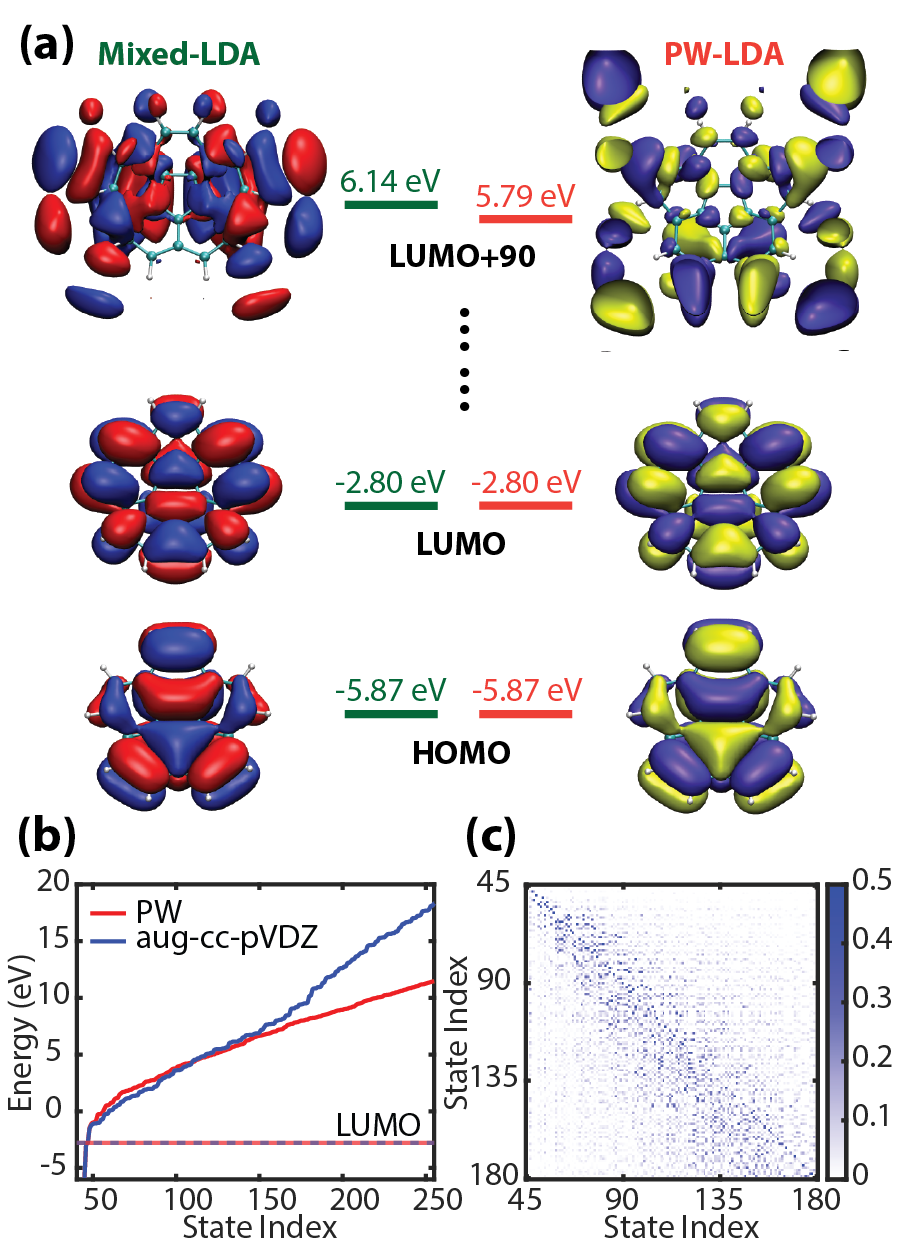}
\caption{\label{fig:fig_2}  \textbf{(a)} HOMO, LUMO, and LUMO+90 densities and energies for C$_{20}$H$_{10}$ calculated at the LDA level with the mixed PW/AO@aug-cc-pVDZ basis (left) and pure PW basis (right). \textbf{(b)} MO energy comparison for C$_{20}$H$_{10}$ for unoccupied states between PW and AO@aug-cc-pVDZ basis at LDA level (HOMO energies have been aligned with a rigid energy-shift of $\Delta\varepsilon=-1.24$ eV). \textbf{(c)} Overlap intensity between unoccupied LDA MOs from PW and AO@aug-cc-pVDZ basis for C$_{20}$H$_{10}$.}
\end{figure}

Fig. \ref{fig:fig_2} presents an orbital analysis of C$_{20}$H$_{10}$ to highlight the key similarities and differences between mixed-basis and pure PW MOs. Near the Fermi level, MOs such as the HOMO and LUMO exhibit comparable spatial density distributions and MO energies across both representations. However, for high-energy unoccupied states, the mixed-basis MOs differ substantially from their PW counterparts. For example, as illustrated in Fig. \ref{fig:fig_2}(a), the LUMO+90 orbital in C$_{20}$H$_{10}$ exhibits more spatially contracted density in the mixed basis due to the AO component, leading to a higher orbital energy. In contrast, the same MO in the pure PW basis is more delocalized and energetically lower. 

The MO energies for unoccupied states derived from the PW and AO bases (after aligning the HOMO energies) are shown in Fig.~\ref{fig:fig_2}(b). As is well known, AO-basis MOs tend to be more spatially localized at higher energies compared to their PW counterparts. This increased localization leads to stronger electron–electron repulsion, and consequently, higher MO energies. This trend is also evident from the overlap matrix between the two sets of orbitals, defined as $S_{ij} = \langle \phi_i^{AO} | \phi_j^{PW} \rangle$, shown in Fig.~\ref{fig:fig_2}(c). At higher energies, the spatial overlap between AO and PW MOs becomes noticeably smaller, reflecting their increasing dissimilarity.

\begin{table}[H]
\centering
\begin{tabular}{c|cc|cc}
\hline
\textbf{System}  & PW-LDA & AO-LDA & PW-Hybrid & Mixed-Hybrid \\ \hline
C$_{20}$H$_{10}$ & 3.05   & 3.07   & 6.71   & 6.70 \\
Chla             & 1.42   & 1.43   & 3.90   & 3.89 \\
ICG-7            & 1.19   & 1.20   & 3.89   & 3.89 \\
Flav-9           & 0.90   & 0.90   & 3.05   & 3.04 \\
\hline
\end{tabular}
\caption{Static LUMO-HOMO gaps (in eV) for four test systems from PW-LDA-DFT, AO@aug-cc-pVDZ LDA-DFT, PW- and mixed PW/AO sparse-stochastic hybrid-DFT with CAM-LDA0 functional.}
\label{tab:static_gap}
\end{table}

Static LDA-DFT and sparse-stochastic hybrid-DFT LUMO–HOMO gaps are presented in Table \ref{tab:static_gap}. Both the initial PW-LDA-DFT and AO-LDA-DFT calculations yield nearly identical results, with differences under 0.02 eV across all test systems. This agreement is maintained at the hybrid-DFT stage. The consistency of ground-state properties establishes a robust foundation for subsequent excited-state calculations using the mixed-basis approach.

\begin{figure}[H]
\centering
\includegraphics[width=3in]{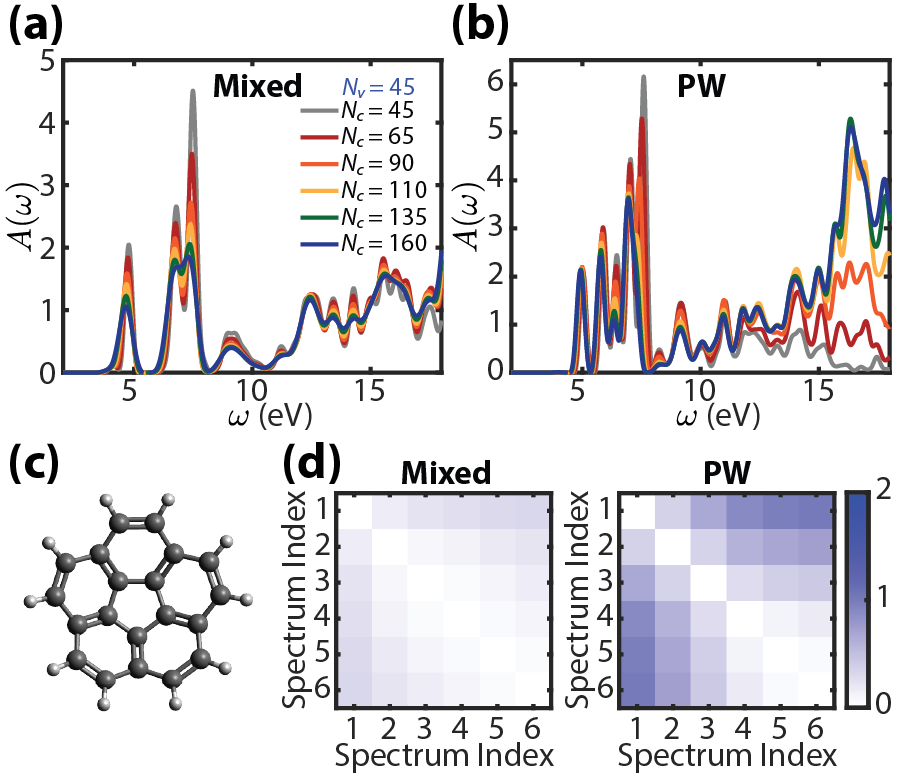}
\caption{\label{fig:fig_4} \textbf{C$_{20}$H$_{10}$}: sparse-stochastic hybrid TDDFT@CAM-LDA0 spectra calculated with different $N_c$ values via \textbf{(a)} mixed PW/AO and \textbf{(b)} PW basis, $N_v=N_{\text{occ}}=45$ is used for all calculations. \textbf{(c)} Molecular structure \textbf{(d)} the spectral deviation matrix $D$ from mixed PW/AO (left) and PW (right) basis.}  
\end{figure}

\begin{figure}[H]
\centering
\includegraphics[width=3in]{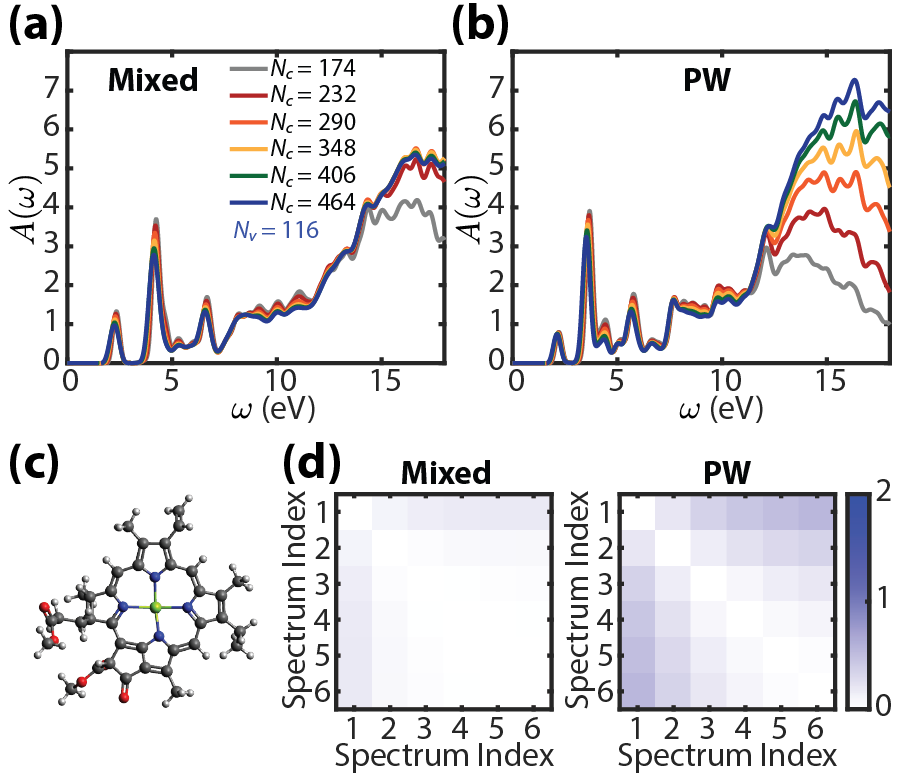}
\caption{\label{fig:fig_5} \textbf{Chla}: sparse-stochastic hybrid TDDFT@CAM-LDA0 spectra calculated with different $N_c$ values via \textbf{(a)} mixed PW/AO and \textbf{(b)} PW basis, $N_v=116$ is used for all calculations. \textbf{(c)} Molecular structure \textbf{(d)} the spectral deviation matrix $D$ from mixed PW/AO (left) and PW (right) basis.}
\end{figure}

\begin{figure}[H]
\centering
\includegraphics[width=3in]{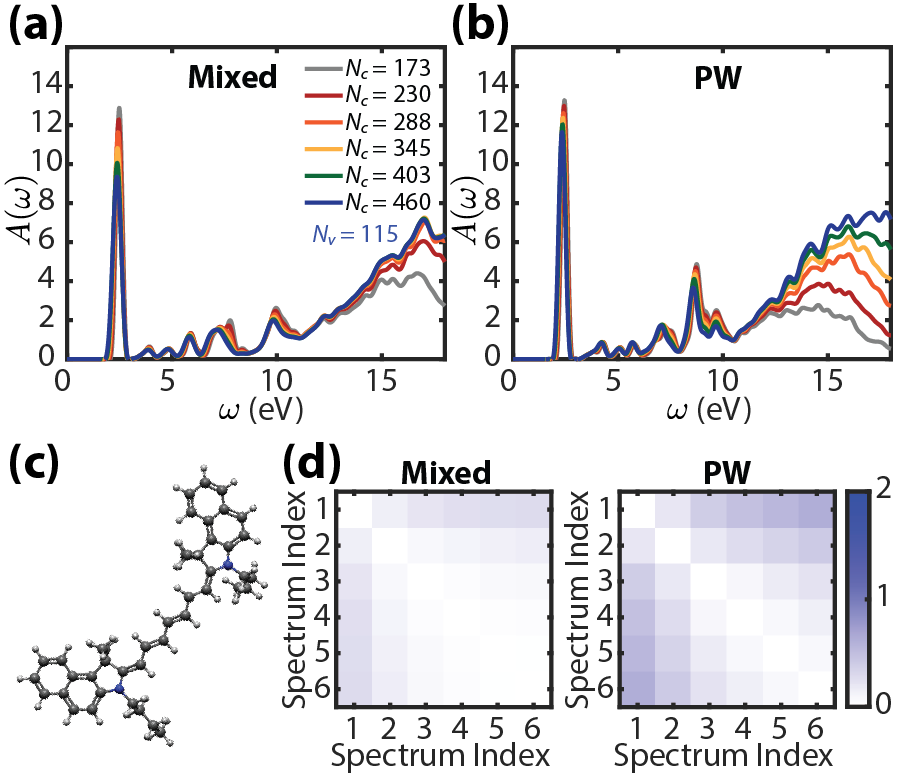}
\caption{\label{fig:fig_6} \textbf{ICG-7}: sparse-stochastic hybrid TDDFT@CAM-LDA0 spectra calculated with different $N_c$ values via \textbf{(a)} mixed PW/AO and \textbf{(b)} PW basis, $N_v=115$ is used for all calculations. \textbf{(c)} Molecular structure \textbf{(d)} the spectral deviation matrix $D$ from mixed PW/AO (left) and PW (right) basis.}
\end{figure}

\begin{figure}[H]
\centering
\includegraphics[width=3in]{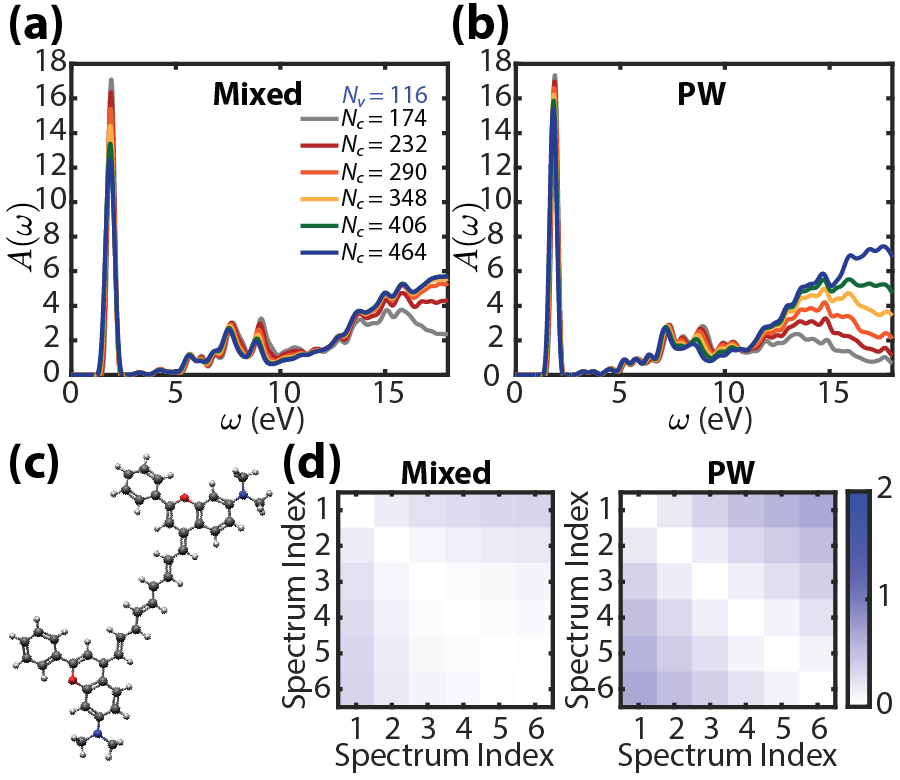}
\caption{\label{fig:fig_7}  \textbf{Flav-9}: sparse-stochastic hybrid TDDFT@CAM-LDA0 spectra calculated with different $N_c$ values via \textbf{(a)} mixed PW/AO and \textbf{(b)} PW basis, $N_v=116$ is used for all calculations. \textbf{(c)} Molecular structure \textbf{(d)} the spectral deviation matrix $D$ from mixed PW/AO (left) and PW (right) basis.}
\end{figure}

Fig. \ref{fig:fig_4} shows sparse-stochastic hybrid TDDFT@CAM-LDA0 spectra for C$_{20}$H$_{10}$ with $N_v=N_{\text{occ}}=45$ calculated with $N_c$ values ranging from 45 to 160, using the mixed PW/AO basis (Fig. \ref{fig:fig_4}(a)) and the pure PW basis (Fig. \ref{fig:fig_4}(b)). Fig. \ref{fig:fig_4}(c) shows the molecular structure, and Fig. \ref{fig:fig_4}(d) displays the spectral deviation matrix $D$ comparing mixed PW/AO (left) and PW (right) bases. Fig. \ref{fig:fig_5} presents similar data for Chla, with $N_v=116$ and $N_c=174-464$; Fig. \ref{fig:fig_6} reports the results for ICG-7, using $N_v=115$ and $N_c=173-460$; finally, Fig. \ref{fig:fig_7} shows results for Flav-9, using $N_v=116$ and $N_c=174-464$, with the same panel arrangements.

Notably, the mixed-basis approach exhibits significantly improved convergence across all four test systems, particularly in the high-energy regions of the spectra. This improvement arises because the more contracted AO basis renders the high-energy MOs more localized, and consequently, the corresponding transitions become more spatially confined.  The spectral deviation matrices, presented in Panel (d) of Figs. \ref{fig:fig_4} through \ref{fig:fig_7}, provide additional insights into spectral behavior during the convergence process. The spectrum index, $i$, ranges from 1 to 6, where each $i$ labels a different value of $N_c$; 1 refers to the smallest $N_c$ used and 6 labels the largest. These matrices consistently show smaller off-diagonal components for mixed-basis calculations compared to pure PW results, indicating more stable and systematic convergence as $N_c$ increases. 

\begin{figure}[H]
\centering
\includegraphics[width=3in]{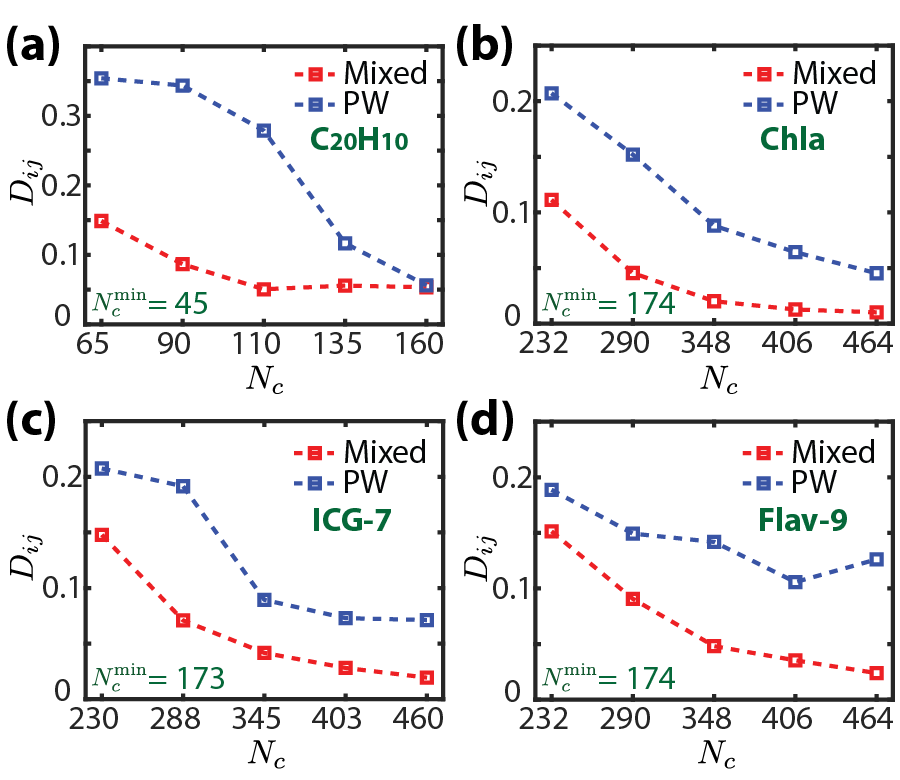}
\caption{\label{fig:fig_8} Deviation between consecutive pairs of spectra for \textbf{(a)} C$_{20}$H$_{10}$, \textbf{(b)} Chla, \textbf{(c)} ICG-7, and \textbf{(d)} Flav-9, comparing the mixed- and PW-basis approaches. Each data point represents the deviation between spectra at a given $N_c$ (shown on the $x$-axis) and the preceding $N_c$ value. The left-most point corresponds to the deviation between the smallest $N_c$ used ($N_c^{\text{min}}$, labeled in each plot) and its immediate next value.}
\end{figure}

To better understand the trend in spectral convergence, we plot the deviation between consecutive spectral pairs, $D_{i,i+1}$, in Fig. \ref{fig:fig_8}. For C$_{20}$H$_{10}$, $N_c^{Mixed}=110=2.5N_v$ achieves equivalent convergence to $N_c^{PW}=160=3.6N_v$ (Fig. \ref{fig:fig_8}(a)). For Chla, $N_c^{Mixed}=290=2.5N_v$ reproduces $N_c^{PW}=494=4.3N_v$ (Fig. \ref{fig:fig_8}(b)). The ICG-7 system follows this trend, where $N_c^{Mixed}=288=2.5N_v$ delivers comparable accuracy to $N_c^{PW}=460=4N_v$ (Fig. \ref{fig:fig_8}(c)). Finally, Flav-9 requires only $N_c^{Mixed}=290=2.5N_v$ to reach the convergence level of $N_c^{PW}=464=4N_v$ (Fig. \ref{fig:fig_8}(d)).

The overall convergence metrics $\delta$ and $\bar{\delta}$ are summarized in Table \ref{tab:deltas}. The mixed-basis calculations show significantly lower $\delta$ and $\bar{\delta}$ values compared to their PW counterparts. Specifically, we observe a $2-3$ fold reduction in the values for $\delta$ and $\bar{\delta}$ using the mixed-based representation of the MOs relative to a purely PW-based representation. This enhanced performance originates from the more compact and chemically intuitive representation of high-energy conduction states in the AO basis, which effectively reduces the number of conduction states ($N_c$) required for achieving spectral convergence.

\begin{table}[H]
\centering
\begin{tabular}{c|cc|cc}
\hline
\textbf{System}  & $\delta$ (Mixed) & $\delta$ (PW) & $\bar{\delta}$ (Mixed) & $\bar{\delta}$ (PW) \\
\hline
C$_{20}$H$_{10}$ &  0.08       &   0.23     &  0.16     &  0.51 \\
Chla             &  0.04       &   0.11     &  0.08     &  0.25 \\
ICG-7            &  0.06       &   0.13     &  0.13     &  0.29 \\
Flav-9           &  0.07       &   0.14     &  0.15     &  0.32 \\
\hline
\end{tabular}
\caption{The average deviation between successive spectra ($\delta$) and the average deviation over all pairs ($\bar{\delta}$) via PW and mixed-basis for four test systems with six tested values of $N_c$, ranging from $N_c=N_v$ to $N_c=4N_v$.}
\label{tab:deltas}
\end{table}

\blue{We note that, while direct localization of PW virtual orbitals could offer an alternative, our mixed-basis approach ensures transferable accuracy without the need for state-specific tuning.} However, our analysis also reveals an important subtlety regarding the first excitation energy, i.e., the optical gap. While the overall spectral shapes show excellent agreement between the PW and mixed-basis calculations, we consistently observe a small ($\sim$ 0.08 eV) but systematic blueshift in the optical gaps.\cite{Bruneval2008,Rocca2008} This observation suggests that, although the mixed basis is highly effective for describing higher-energy excitations and provides significantly faster spectral convergence, special care is required when precise determination of the optical gap is needed.

\section{Conclusion}
This study demonstrates the advantages of combining PW and AO basis sets within the sparse-stochastic hybrid TDDFT framework, offering a balanced approach that maintains accuracy while improving computational efficiency. The mixed-basis method, leveraging PW-derived occupied MOs and carefully selected AO-derived unoccupied MOs, achieves results with superior spectral convergence with respect to the number of conduction states. Our analysis confirms excellent agreement for ground-state properties in sparse-stochastic hybrid DFT calculations, where the LUMO-HOMO gap deviations between the PW and mixed-basis calculations are below 0.02 eV. The sparse-stochastic hybrid TDDFT spectral calculations show that the localized nature of AOs enhances convergence for high-energy excitations and offers an overall $2-3$ folds faster convergence. \blue{We also acknowledge that while the current $\sim25\%$ PW retaining ratio in the unoccupied manifold effectively mitigates blueshifts in neutral or positively charged organic systems, its extension to charged species or transition metal complexes will require further system-dependent adjustments of the mixing composition.}

Looking ahead, this work opens several promising avenues for development, including optimized mixing ratios of PW- and AO-derived MOs in the unoccupied space, extension to periodic boundary conditions, and integration with fragment-based methods for large-scale simulations. 

\begin{acknowledgement}
    This work is supported  by the National Science Foundation (NSF) under Grant No. CHE-2245253. Computational resources for simulations were provided by both the Expanse cluster at the San Diego Supercomputer Center through allocation CHE-230099 under the Advanced Cyberinfrastructure Coordination
    Ecosystem: Services \& Support (ACCESS) program and the Hoffman2 Shared Cluster provided by UCLA Office of Advanced Research Computing’s Research Technology Group.
\end{acknowledgement}

\section*{Appendix A: Basis-Set Benchmarking}
\begin{figure}[H]
\centering
\includegraphics[width=3in]{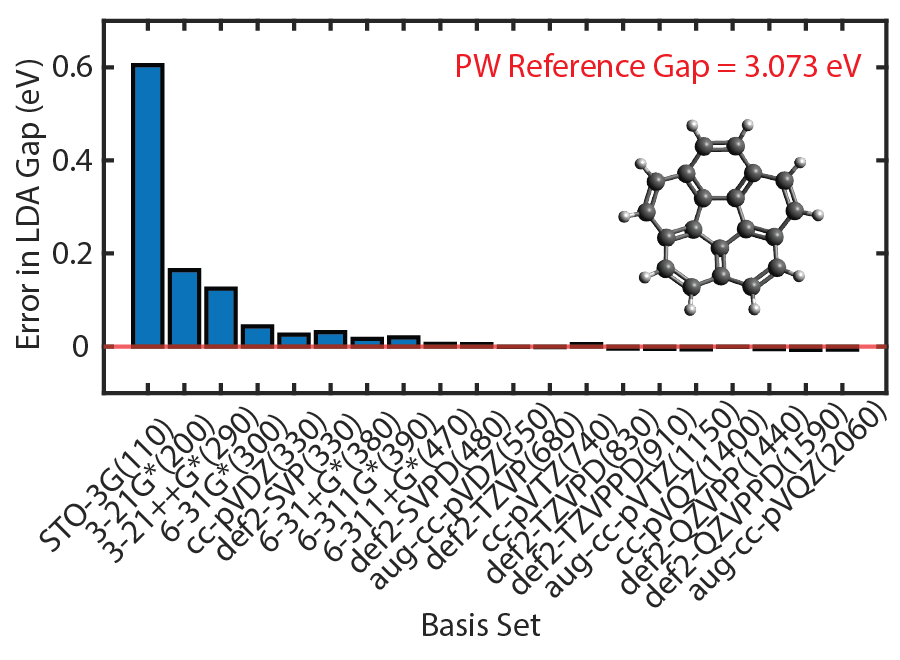}
\caption{\label{fig:fig_basis} LUMO-HOMO LDA gap differences (compared to the PW-LDA result) in eV calculated for C$_{20}$H$_{10}$, using various AO basis sets. The numbers in parentheses indicate the total number of basis functions used. The minimum deviation AO set is aug-cc-pVDZ with 550 total basis functions, where the gap difference between AO and PW is $-6.05\times10^{-4}$ eV.}
\end{figure}

Localized AO basis sets vary widely in their composition and completeness, ranging from minimal sets like STO-3G to highly extensive ones such as aug-cc-pVQZ. In principle, as an AO basis set becomes more complete, its results should asymptotically approach those of PW calculations. However, this increased completeness comes at a significantly higher computational cost. To identify an optimal AO basis set for the mixed-basis scheme, we benchmark several AO basis sets by performing LDA-DFT calculations and comparing the resulting LUMO-HOMO gaps against those from a reference PW LDA-DFT calculation. The error is defined as the difference between AO and the reference PW gaps. All benchmarks are conducted on the C$_{20}$H$_{10}$ (Fig. \ref{fig:fig_basis}). \blue{Table \ref{tab:c20_ngh_basis} demonstrates that static gaps (calculated via our PW/AO sparse-stochastic hybrid-DFT with CAM-LDA0 functional) of C$_{20}$H$_{10}$ exhibit minimal sensitivity ($\leq0.037$ eV variation) across nine different AO basis sets. The aug-cc-pVDZ basis performs particularly well, differing by only 0.009 eV from the PW reference.}

\begin{table}[H]
\centering
\begin{tabular}{c|c|c}
\hline
\textbf{Basis Set}  & \textbf{Static Gap (eV)} & \textbf{Abs. Difference (eV)} \\ \hline
Pure PW        & 6.712       & (Reference)         \\
aug-cc-pVDZ    & 6.704       & 0.009         \\
aug-cc-pVTZ    & 6.703       & 0.009         \\
aug-cc-pVQZ    & 6.701       & 0.012         \\
3-21G*         & 6.694       & 0.018         \\
cc-pVDZ        & 6.680       & 0.032         \\
def2-SVP       & 6.678       & 0.034         \\
STO-3G         & 6.677       & 0.035         \\
6-311G*        & 6.675       & 0.037         \\
\hline
\end{tabular}
\caption{\blue{Static LUMO-HOMO gap (in eV) of C$_{20}$H$_{10}$ calculated via PW/AO sparse-stochastic hybrid-DFT with CAM-LDA0 functional for various AO basis sets compared to the pure PW reference.}}
\label{tab:c20_ngh_basis}
\end{table}

Since our calculations employ real-space grids, we require basis functions with smooth spatial decay to maintain numerical stability in integrations.\cite{Becke1988,Becke1988pra} The aug-cc-pVDZ basis set satisfies this need by avoiding excessively sharp Gaussian exponents ($\zeta$): carbon's tightest exponent in aug-cc-pVDZ has $\zeta\sim8\times10^3$ Bohr$^{-2}$ is far smoother than in sharper sets like 6-31G$^*$ with $\zeta\sim2\times10^4$ Bohr$^{-2}$ or STO-3G with $\zeta\sim7\times10^4$ Bohr$^{-2}$.\cite{Dunning1989,Pritchard2019} This careful balance prevents the grid artifacts that would arise from extreme exponents while maintaining accuracy for both core and valence electrons.

\section*{Appendix B: Comparison of Orthonormalization Techniques}
The orthonormality condition on the numerical grid is checked via the Frobenius error
\begin{equation}
    \|I - S\|_2 = \sqrt{\sum_{i,j}|(I - S)_{ij}|^2},
\end{equation}
where $I$ is the identity matrix and $S$ is the overlap matrix of the orthonormalized MOs (Table \ref{tab:frobenius}).
\begin{table}[H]  
\centering  
    \begin{tabular}{c|ccc}  
    \hline
        & Householder& Gram-Schmidt& Löwdin\\  
        \hline  
        C$_{20}$H$_{10}$& \makebox[2cm]{$1.41\times10^{-14}$} & \makebox[2cm]{$1.63\times10^{-14}$} & \makebox[2cm]{$2.05\times10^{-13}$}\\  
        Chla& \makebox[2cm]{$4.10\times10^{-14}$}& \makebox[2cm]{$3.96\times10^{-14}$}& \makebox[2cm]{$2.06\times10^{-13}$}\\  
        ICG-7& \makebox[2cm]{$4.24\times10^{-14}$}& \makebox[2cm]{$4.24\times10^{-14}$}& \makebox[2cm]{$4.64\times10^{-13}$}\\  
        Flav-9& \makebox[2cm]{$4.72\times10^{-14}$}& \makebox[2cm]{$4.58\times10^{-14}$}& \makebox[2cm]{$3.64\times10^{-13}$}\\  
        \hline
    \end{tabular}  
\caption{\label{tab:frobenius} $\|{I - S}\|_2$ Frobenius error using different orthonormalization procedures.}  
\end{table}
To identify the optimal procedure here, we benchmark results from three different approaches: Löwdin orthonormalization, Gram-Schmidt, and Householder transformation. Once the cost of making $S$ is included, all methods scale as 
$\mathcal{O}(N_gM^2)$, where $M=N_v+N_c$ is the total number of MOs and $N_g$ is the number of grid-points. Although Gram-Schmidt offers slightly better accuracy for larger molecules, we use the Householder transformation due to its availability as highly optimized low-level code in Python packages. 

\section*{Appendix C: \blue{Real-space Grid and} Computational Parameters}
General computational parameters are tabulated in Table \ref{tab:grid_params}.
\begin{table}[H]
\centering
\begin{tabular}{c|ccc|c|cc|c}
\hline
\textbf{System} & $N_x$ & $N_y$ & $N_z$ & $dx$ & $N_v$ & $N_c$ & $N_{k_{\text{low}}}$\\ \hline
C$_{20}$H$_{10}$ &70  &70  &40  &0.4  &45  &$45-160$  &201   \\
Chla             &106 &96  &80  &0.4  &116  &$174-464$  &867   \\
ICG-7            &130 &100 &80  &0.4  &115  &$173-460$  &1127  \\
Flav-9           &156 &100 &70  &0.4  &116  &$174-464$  &1195  \\
\hline
\end{tabular}
\caption{Computational parameters for the four test systems: grid extents, isotropic grid-spacing of $dx=dy=dz=0.4$ Bohr is used for all systems, $N_v$ values, and $N_c$ ranges used for sparse-stochastic hybrid DFT and TDDFT calculations. $N_{k_{\text{low}}}$ is the number of deterministically treated long-wavelength terms in the exchange kernel $u(k)$. The high-$k$ space is represented with 1000 sparse-stochastic vectors, details are provided in Refs. \cite{bradbury_neargap_2023,Sereda2024}.}
\label{tab:grid_params}
\end{table}

\begin{figure}[H]
\centering
\includegraphics[width=3in]{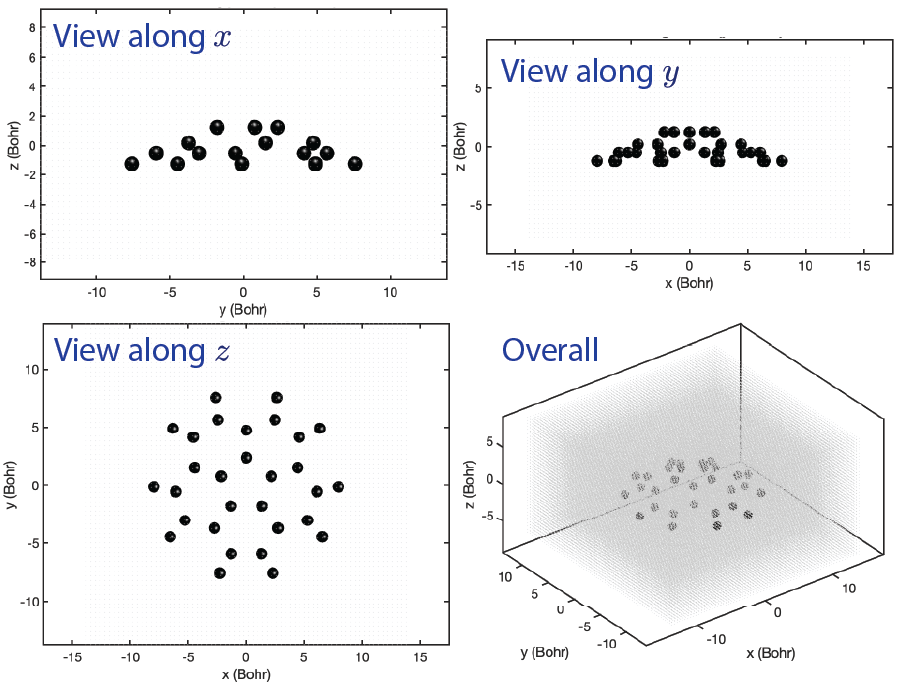}
\caption{\label{fig:grids} \blue{C$_{20}$H$_{10}$ molecule (black balls) and the overall simulation grid (gray dots) viewed from different directions.}}
\end{figure}

\blue{Fig. \ref{fig:grids} illustrates an example for the real-space grid (``box") configuration for the C$_{20}$H$_{10}$ molecule, employing a uniform grid spacing of 0.4 Bohr. The molecule is centered at the origin, and the grid extends along each Cartesian direction with a minimum 6 Bohr padding to ensure negligible boundary effects. Identical criteria (fixed spacing + padding) are applied to other molecules, with box sizes adjusted to accommodate their respective dimensions.}

\bibliography{acssamp}

\providecommand{\latin}[1]{#1}
\makeatletter
\providecommand{\doi}
  {\begingroup\let\do\@makeother\dospecials
  \catcode`\{=1 \catcode`\}=2 \doi@aux}
\providecommand{\doi@aux}[1]{\endgroup\texttt{#1}}
\makeatother
\providecommand*\mcitethebibliography{\thebibliography}
\csname @ifundefined\endcsname{endmcitethebibliography}  {\let\endmcitethebibliography\endthebibliography}{}
\begin{mcitethebibliography}{59}
\providecommand*\natexlab[1]{#1}
\providecommand*\mciteSetBstSublistMode[1]{}
\providecommand*\mciteSetBstMaxWidthForm[2]{}
\providecommand*\mciteBstWouldAddEndPuncttrue
  {\def\EndOfBibitem{\unskip.}}
\providecommand*\mciteBstWouldAddEndPunctfalse
  {\let\EndOfBibitem\relax}
\providecommand*\mciteSetBstMidEndSepPunct[3]{}
\providecommand*\mciteSetBstSublistLabelBeginEnd[3]{}
\providecommand*\EndOfBibitem{}
\mciteSetBstSublistMode{f}
\mciteSetBstMaxWidthForm{subitem}{(\alph{mcitesubitemcount})}
\mciteSetBstSublistLabelBeginEnd
  {\mcitemaxwidthsubitemform\space}
  {\relax}
  {\relax}

\bibitem[Runge and Gross(1984)Runge, and Gross]{RungeGross1984}
Runge,~E.; Gross,~E. K.~U. Density-Functional Theory for Time-Dependent Systems. \emph{Phys. Rev. Lett.} \textbf{1984}, \emph{52}, 997--1000\relax
\mciteBstWouldAddEndPuncttrue
\mciteSetBstMidEndSepPunct{\mcitedefaultmidpunct}
{\mcitedefaultendpunct}{\mcitedefaultseppunct}\relax
\EndOfBibitem
\bibitem[Casida(1995)]{casida1995}
Casida,~M.~E. Time-Dependent Density Functional Response Theory for Molecules. \emph{World Sci.} \textbf{1995}, \emph{1}, 155--192\relax
\mciteBstWouldAddEndPuncttrue
\mciteSetBstMidEndSepPunct{\mcitedefaultmidpunct}
{\mcitedefaultendpunct}{\mcitedefaultseppunct}\relax
\EndOfBibitem
\bibitem[Jamorski \latin{et~al.}(1996)Jamorski, Casida, and Salahub]{jamorski1996}
Jamorski,~C.; Casida,~M.~E.; Salahub,~D.~R. Dynamic polarizabilities and excitation spectra from a molecular implementation of time-dependent density-functional response theory: N2 as a case study. \emph{J. Chem. Phys.} \textbf{1996}, \emph{104}, 5134--5147\relax
\mciteBstWouldAddEndPuncttrue
\mciteSetBstMidEndSepPunct{\mcitedefaultmidpunct}
{\mcitedefaultendpunct}{\mcitedefaultseppunct}\relax
\EndOfBibitem
\bibitem[Yabana and Bertsch(1997)Yabana, and Bertsch]{yabana1997}
Yabana,~K.; Bertsch,~G. Optical response of small carbon clusters. \emph{Z. Phys. D: At., Mol. Clusters} \textbf{1997}, \emph{42}, 219--225\relax
\mciteBstWouldAddEndPuncttrue
\mciteSetBstMidEndSepPunct{\mcitedefaultmidpunct}
{\mcitedefaultendpunct}{\mcitedefaultseppunct}\relax
\EndOfBibitem
\bibitem[Yabana and Bertsch(1999)Yabana, and Bertsch]{yabana1999}
Yabana,~K.; Bertsch,~G.~F. Time-Dependent Local-Density Approximation in Real Time: Application to Conjugated Molecules. \emph{Int. J. Quantum Chem.} \textbf{1999}, \emph{75}, 55--66\relax
\mciteBstWouldAddEndPuncttrue
\mciteSetBstMidEndSepPunct{\mcitedefaultmidpunct}
{\mcitedefaultendpunct}{\mcitedefaultseppunct}\relax
\EndOfBibitem
\bibitem[Lopata \latin{et~al.}(2012)Lopata, {Van Kuiken}, Khalil, and Govind]{lopata2012}
Lopata,~K.; {Van Kuiken},~B.~E.; Khalil,~M.; Govind,~N. Linear-Response and Real-Time Time-Dependent Density Functional Theory Studies of Core-Level Near-Edge X-Ray Absorption. \emph{J. Chem. Theory Comput.} \textbf{2012}, \emph{8}, 3284--3292\relax
\mciteBstWouldAddEndPuncttrue
\mciteSetBstMidEndSepPunct{\mcitedefaultmidpunct}
{\mcitedefaultendpunct}{\mcitedefaultseppunct}\relax
\EndOfBibitem
\bibitem[{DeBeer George} \latin{et~al.}(2008){DeBeer George}, Petrenko, and Neese]{debeer2008}
{DeBeer George},~S.; Petrenko,~T.; Neese,~F. Time-dependent density functional calculations of ligand K-edge X-ray absorption spectra. \emph{Inorg. Chim. Acta} \textbf{2008}, \emph{361}, 965--972\relax
\mciteBstWouldAddEndPuncttrue
\mciteSetBstMidEndSepPunct{\mcitedefaultmidpunct}
{\mcitedefaultendpunct}{\mcitedefaultseppunct}\relax
\EndOfBibitem
\bibitem[Lerm\'{e} \latin{et~al.}(1998)Lerm\'{e}, Palpant, Pr\'{e}vel, Cottancin, Pellarin, Treilleux, Vialle, Perez, and Broyer]{lerme1998}
Lerm\'{e},~J.; Palpant,~B.; Pr\'{e}vel,~B.; Cottancin,~E.; Pellarin,~M.; Treilleux,~M.; Vialle,~J.; Perez,~A.; Broyer,~M. Optical properties of gold metal clusters: A time-dependent local-density-approximation investigation. \emph{Eur. Phys. J. D} \textbf{1998}, \emph{4}, 95--108\relax
\mciteBstWouldAddEndPuncttrue
\mciteSetBstMidEndSepPunct{\mcitedefaultmidpunct}
{\mcitedefaultendpunct}{\mcitedefaultseppunct}\relax
\EndOfBibitem
\bibitem[Qian \latin{et~al.}(2006)Qian, Li, Lin, and Yip]{qian2006}
Qian,~X.; Li,~J.; Lin,~X.; Yip,~S. Time-dependent density functional theory with ultrasoft pseudopotentials: Real-time electron propagation across a molecular junction. \emph{Phys. Rev. B: Condens. Matter Mater. Phys.} \textbf{2006}, \emph{73}, 035408\relax
\mciteBstWouldAddEndPuncttrue
\mciteSetBstMidEndSepPunct{\mcitedefaultmidpunct}
{\mcitedefaultendpunct}{\mcitedefaultseppunct}\relax
\EndOfBibitem
\bibitem[Baer and Neuhauser(2003)Baer, and Neuhauser]{baer2003}
Baer,~R.; Neuhauser,~D. Ab Initio Electrical Conductance of a Molecular Wire. \emph{Int. J. Quantum Chem.} \textbf{2003}, \emph{91}, 524--532\relax
\mciteBstWouldAddEndPuncttrue
\mciteSetBstMidEndSepPunct{\mcitedefaultmidpunct}
{\mcitedefaultendpunct}{\mcitedefaultseppunct}\relax
\EndOfBibitem
\bibitem[Baer and Gould(2001)Baer, and Gould]{baer2001}
Baer,~R.; Gould,~R. A method for ab initio nonlinear electron-density evolution. \emph{J. Chem. Phys.} \textbf{2001}, \emph{114}, 3385--3392\relax
\mciteBstWouldAddEndPuncttrue
\mciteSetBstMidEndSepPunct{\mcitedefaultmidpunct}
{\mcitedefaultendpunct}{\mcitedefaultseppunct}\relax
\EndOfBibitem
\bibitem[Baer and Neuhauser(2004)Baer, and Neuhauser]{baer2004}
Baer,~R.; Neuhauser,~D. Real-time linear response for time-dependent density-functional theory. \emph{The Journal of Chemical Physics} \textbf{2004}, \emph{121}, 9803--9807\relax
\mciteBstWouldAddEndPuncttrue
\mciteSetBstMidEndSepPunct{\mcitedefaultmidpunct}
{\mcitedefaultendpunct}{\mcitedefaultseppunct}\relax
\EndOfBibitem
\bibitem[Zelovich \latin{et~al.}(2014)Zelovich, Kronik, and Hod]{zelovich2014}
Zelovich,~T.; Kronik,~L.; Hod,~O. State Representation Approach for Atomistic Time-Dependent Transport Calculations in Molecular Junctions. \emph{J. Chem. Theory Comput.} \textbf{2014}, \emph{10}, 2927--2941\relax
\mciteBstWouldAddEndPuncttrue
\mciteSetBstMidEndSepPunct{\mcitedefaultmidpunct}
{\mcitedefaultendpunct}{\mcitedefaultseppunct}\relax
\EndOfBibitem
\bibitem[Ding \latin{et~al.}(2013)Ding, {Van Kuiken}, Eichinger, and Li]{ding2013}
Ding,~F.; {Van Kuiken},~B.~E.; Eichinger,~B.~E.; Li,~X. An efficient method for calculating dynamical hyperpolarizabilities using real-time time-dependent density functional theory. \emph{J. Chem. Phys.} \textbf{2013}, \emph{138}, 064104\relax
\mciteBstWouldAddEndPuncttrue
\mciteSetBstMidEndSepPunct{\mcitedefaultmidpunct}
{\mcitedefaultendpunct}{\mcitedefaultseppunct}\relax
\EndOfBibitem
\bibitem[Becke(1993)]{becke1993}
Becke,~A.~D. Density-functional thermochemistry. {III}. The role of exact exchange. \emph{J. Chem. Phys.} \textbf{1993}, \emph{98}, 5648--5652\relax
\mciteBstWouldAddEndPuncttrue
\mciteSetBstMidEndSepPunct{\mcitedefaultmidpunct}
{\mcitedefaultendpunct}{\mcitedefaultseppunct}\relax
\EndOfBibitem
\bibitem[Baer \latin{et~al.}(2010)Baer, Livshits, and Salzner]{baer2010}
Baer,~R.; Livshits,~E.; Salzner,~U. Tuned Range-Separated Hybrids in Density Functional Theory. \emph{Annu. Rev. Phys. Chem.} \textbf{2010}, \emph{61}, 85--109\relax
\mciteBstWouldAddEndPuncttrue
\mciteSetBstMidEndSepPunct{\mcitedefaultmidpunct}
{\mcitedefaultendpunct}{\mcitedefaultseppunct}\relax
\EndOfBibitem
\bibitem[Leininger \latin{et~al.}(1997)Leininger, Stoll, Werner, and Savin]{leininger1997}
Leininger,~T.; Stoll,~H.; Werner,~H.-J.; Savin,~A. Combining long-range configuration interaction with short-range density functionals. \emph{Chem. Phys. Lett.} \textbf{1997}, \emph{275}, 151--160\relax
\mciteBstWouldAddEndPuncttrue
\mciteSetBstMidEndSepPunct{\mcitedefaultmidpunct}
{\mcitedefaultendpunct}{\mcitedefaultseppunct}\relax
\EndOfBibitem
\bibitem[Yanai \latin{et~al.}(2004)Yanai, Tew, and Handy]{Yanai2004}
Yanai,~T.; Tew,~D.~P.; Handy,~N.~C. A new hybrid exchange–correlation functional using the Coulomb-attenuating method (CAM-B3LYP). \emph{Chemical Physics Letters} \textbf{2004}, \emph{393}, 51–57\relax
\mciteBstWouldAddEndPuncttrue
\mciteSetBstMidEndSepPunct{\mcitedefaultmidpunct}
{\mcitedefaultendpunct}{\mcitedefaultseppunct}\relax
\EndOfBibitem
\bibitem[Karolewski \latin{et~al.}(2011)Karolewski, Stein, Baer, and K{\"u}mmel]{karolewski2011}
Karolewski,~A.; Stein,~T.; Baer,~R.; K{\"u}mmel,~S. Communication: Tailoring the optical gap in light-harvesting molecules. \emph{J. Chem. Phys.} \textbf{2011}, \emph{134}, 151101\relax
\mciteBstWouldAddEndPuncttrue
\mciteSetBstMidEndSepPunct{\mcitedefaultmidpunct}
{\mcitedefaultendpunct}{\mcitedefaultseppunct}\relax
\EndOfBibitem
\bibitem[Kronik \latin{et~al.}(2012)Kronik, Stein, Refaely-Abramson, and Baer]{kronik2012}
Kronik,~L.; Stein,~T.; Refaely-Abramson,~S.; Baer,~R. Excitation Gaps of Finite-Sized Systems from Optimally Tuned Range-Separated Hybrid Functionals. \emph{J. Chem. Theory Comput.} \textbf{2012}, \emph{8}, 1515--1531\relax
\mciteBstWouldAddEndPuncttrue
\mciteSetBstMidEndSepPunct{\mcitedefaultmidpunct}
{\mcitedefaultendpunct}{\mcitedefaultseppunct}\relax
\EndOfBibitem
\bibitem[Vl{\v{c}}ek \latin{et~al.}(2019)Vl{\v{c}}ek, Baer, and Neuhauser]{vlcek2019}
Vl{\v{c}}ek,~V.; Baer,~R.; Neuhauser,~D. Stochastic time-dependent {DFT} with optimally tuned range-separated hybrids: {Application} to excitonic effects in large phosphorene sheets. \emph{J. Chem. Phys.} \textbf{2019}, \emph{150}, 184118\relax
\mciteBstWouldAddEndPuncttrue
\mciteSetBstMidEndSepPunct{\mcitedefaultmidpunct}
{\mcitedefaultendpunct}{\mcitedefaultseppunct}\relax
\EndOfBibitem
\bibitem[Perdew and Wang(1992)Perdew, and Wang]{PerdewWang1992}
Perdew,~J.~P.; Wang,~Y. Accurate and simple analytic representation of the electron-gas correlation energy. \emph{Phys. Rev. B} \textbf{1992}, \emph{45}, 13244--13249\relax
\mciteBstWouldAddEndPuncttrue
\mciteSetBstMidEndSepPunct{\mcitedefaultmidpunct}
{\mcitedefaultendpunct}{\mcitedefaultseppunct}\relax
\EndOfBibitem
\bibitem[Perdew \latin{et~al.}(1996)Perdew, Burke, and Ernzerhof]{perdew_generalized_1996}
Perdew,~J.~P.; Burke,~K.; Ernzerhof,~M. Generalized {Gradient} {Approximation} {Made} {Simple}. \emph{Physical Review Letters} \textbf{1996}, \emph{77}, 3865--3868\relax
\mciteBstWouldAddEndPuncttrue
\mciteSetBstMidEndSepPunct{\mcitedefaultmidpunct}
{\mcitedefaultendpunct}{\mcitedefaultseppunct}\relax
\EndOfBibitem
\bibitem[Bradbury \latin{et~al.}(2023)Bradbury, Allen, Nguyen, and Neuhauser]{bradbury_neargap_2023}
Bradbury,~N.~C.; Allen,~T.; Nguyen,~M.; Neuhauser,~D. Deterministic/Fragmented-Stochastic Exchange for Large-Scale Hybrid DFT Calculations. \emph{Journal of Chemical Theory and Computation} \textbf{2023}, \emph{19}, 9239–9247\relax
\mciteBstWouldAddEndPuncttrue
\mciteSetBstMidEndSepPunct{\mcitedefaultmidpunct}
{\mcitedefaultendpunct}{\mcitedefaultseppunct}\relax
\EndOfBibitem
\bibitem[Sereda \latin{et~al.}(2024)Sereda, Allen, Bradbury, Ibrahim, and Neuhauser]{Sereda2024}
Sereda,~M.; Allen,~T.; Bradbury,~N.~C.; Ibrahim,~K.~Z.; Neuhauser,~D. Sparse-Stochastic Fragmented Exchange for Large-Scale Hybrid Time-Dependent Density Functional Theory Calculations. \emph{Journal of Chemical Theory and Computation} \textbf{2024}, \emph{20}, 4196–4204\relax
\mciteBstWouldAddEndPuncttrue
\mciteSetBstMidEndSepPunct{\mcitedefaultmidpunct}
{\mcitedefaultendpunct}{\mcitedefaultseppunct}\relax
\EndOfBibitem
\bibitem[Booth \latin{et~al.}(2016)Booth, Tsatsoulis, Chan, and Grüneis]{Booth2016}
Booth,~G.~H.; Tsatsoulis,~T.; Chan,~G. K.~L.; Grüneis,~A. From plane waves to local Gaussians for the simulation of correlated periodic systems. \emph{Journal of Chemical Physics} \textbf{2016}, \emph{145}\relax
\mciteBstWouldAddEndPuncttrue
\mciteSetBstMidEndSepPunct{\mcitedefaultmidpunct}
{\mcitedefaultendpunct}{\mcitedefaultseppunct}\relax
\EndOfBibitem
\bibitem[Sun \latin{et~al.}(2017)Sun, Berkelbach, McClain, and Chan]{Sun2017}
Sun,~Q.; Berkelbach,~T.~C.; McClain,~J.~D.; Chan,~G. K.~L. Gaussian and plane-wave mixed density fitting for periodic systems. \emph{Journal of Chemical Physics} \textbf{2017}, \emph{147}\relax
\mciteBstWouldAddEndPuncttrue
\mciteSetBstMidEndSepPunct{\mcitedefaultmidpunct}
{\mcitedefaultendpunct}{\mcitedefaultseppunct}\relax
\EndOfBibitem
\bibitem[Colle \latin{et~al.}(1987)Colle, Fortunell, and Simonucci]{Colle1987}
Colle,~R.; Fortunell,~A.; Simonucci,~S. A Mixed Basis Set of Plane Waves and Hermite Gaussian Functions. Analytic Expressions of Prototype Integrals. \emph{IL NUOVO CIMENTO} \textbf{1987}, \emph{9}, 969--977\relax
\mciteBstWouldAddEndPuncttrue
\mciteSetBstMidEndSepPunct{\mcitedefaultmidpunct}
{\mcitedefaultendpunct}{\mcitedefaultseppunct}\relax
\EndOfBibitem
\bibitem[Euwema(1971)]{Euwema1971}
Euwema,~R.~N. Plane-wave-Gaussian energy-band study of Nb. \emph{Physical Review B} \textbf{1971}, \emph{4}, 4332--4341\relax
\mciteBstWouldAddEndPuncttrue
\mciteSetBstMidEndSepPunct{\mcitedefaultmidpunct}
{\mcitedefaultendpunct}{\mcitedefaultseppunct}\relax
\EndOfBibitem
\bibitem[Fabian \latin{et~al.}(2025)Fabian, Rabani, and Baer]{Fabian2025}
Fabian,~M.~D.; Rabani,~E.; Baer,~R. Compact Gaussian basis sets for stochastic DFT calculations. \emph{Chemical Physics Letters} \textbf{2025}, \emph{865}\relax
\mciteBstWouldAddEndPuncttrue
\mciteSetBstMidEndSepPunct{\mcitedefaultmidpunct}
{\mcitedefaultendpunct}{\mcitedefaultseppunct}\relax
\EndOfBibitem
\bibitem[Cársky \latin{et~al.}(1998)Cársky, Polášek, and Heyrovsk´yheyrovsk´y]{Carsky1998}
Cársky,~P.; Polášek,~M.; Heyrovsk´yheyrovsk´y,~J. Evaluation of Molecular Integrals in a Mixed Gaussian and Plane-Wave Basis by Rys Quadrature. \emph{JOURNAL OF COMPUTATIONAL PHYSICS} \textbf{1998}, \emph{143}, 266--277\relax
\mciteBstWouldAddEndPuncttrue
\mciteSetBstMidEndSepPunct{\mcitedefaultmidpunct}
{\mcitedefaultendpunct}{\mcitedefaultseppunct}\relax
\EndOfBibitem
\bibitem[Liu and Herbert(2015)Liu, and Herbert]{liu2015}
Liu,~J.; Herbert,~J. An efficient and accurate approximation to time-dependent density functional theory for systems of weakly coupled monomers. \emph{J. Chem. Phys.} \textbf{2015}, \emph{143}, 034106\relax
\mciteBstWouldAddEndPuncttrue
\mciteSetBstMidEndSepPunct{\mcitedefaultmidpunct}
{\mcitedefaultendpunct}{\mcitedefaultseppunct}\relax
\EndOfBibitem
\bibitem[Ko \latin{et~al.}(2021)Ko, Santra, and {DiStasio Jr}]{ko2021}
Ko,~H.-Y.; Santra,~B.; {DiStasio Jr},~R.~A. Enabling Large-Scale Condensed-Phase Hybrid Density Functional Theory-Based Ab Initio Molecular Dynamics II: Extensions to the Isobaric–Isoenthalpic and Isobaric–Isothermal Ensembles. \emph{J. Chem. Theory Comput.} \textbf{2021}, \emph{17}, 7789--7813\relax
\mciteBstWouldAddEndPuncttrue
\mciteSetBstMidEndSepPunct{\mcitedefaultmidpunct}
{\mcitedefaultendpunct}{\mcitedefaultseppunct}\relax
\EndOfBibitem
\bibitem[Samsonidze \latin{et~al.}(2011)Samsonidze, Jain, Deslippe, Cohen, and Louie]{Samsonidze2011}
Samsonidze,~G.; Jain,~M.; Deslippe,~J.; Cohen,~M.~L.; Louie,~S.~G. Simple approximate physical orbitals for GW quasiparticle calculations. \emph{Physical Review Letters} \textbf{2011}, \emph{107}\relax
\mciteBstWouldAddEndPuncttrue
\mciteSetBstMidEndSepPunct{\mcitedefaultmidpunct}
{\mcitedefaultendpunct}{\mcitedefaultseppunct}\relax
\EndOfBibitem
\bibitem[Baroni \latin{et~al.}(1987)Baroni, Giannozzi, and Testa]{Baroni1987}
Baroni,~S.; Giannozzi,~P.; Testa,~A. Green's-Function Approach to Linear Response in Solids. \emph{Phys. Rev. Lett.} \textbf{1987}, \emph{58}, 1861--1864\relax
\mciteBstWouldAddEndPuncttrue
\mciteSetBstMidEndSepPunct{\mcitedefaultmidpunct}
{\mcitedefaultendpunct}{\mcitedefaultseppunct}\relax
\EndOfBibitem
\bibitem[Leininger \latin{et~al.}(1997)Leininger, Stoll, Werner, and Savin]{savin1997}
Leininger,~T.; Stoll,~H.; Werner,~H.-J.; Savin,~A. Combining long-range configuration interaction with short-range density functionals. \emph{Chemical Physics Letters} \textbf{1997}, \emph{275}, 151--160\relax
\mciteBstWouldAddEndPuncttrue
\mciteSetBstMidEndSepPunct{\mcitedefaultmidpunct}
{\mcitedefaultendpunct}{\mcitedefaultseppunct}\relax
\EndOfBibitem
\bibitem[Bradbury \latin{et~al.}(2022)Bradbury, Nguyen, Caram, and Neuhauser]{bradbury_bethesalpeter_2022}
Bradbury,~N.~C.; Nguyen,~M.; Caram,~J.~R.; Neuhauser,~D. Bethe–{Salpeter} equation spectra for very large systems. \emph{The Journal of Chemical Physics} \textbf{2022}, \emph{157}, 031104\relax
\mciteBstWouldAddEndPuncttrue
\mciteSetBstMidEndSepPunct{\mcitedefaultmidpunct}
{\mcitedefaultendpunct}{\mcitedefaultseppunct}\relax
\EndOfBibitem
\bibitem[Bradbury \latin{et~al.}(2023)Bradbury, Allen, Nguyen, Ibrahim, and Neuhauser]{bradbury_optimized_2023}
Bradbury,~N.~C.; Allen,~T.; Nguyen,~M.; Ibrahim,~K.~Z.; Neuhauser,~D. Optimized attenuated interaction: {Enabling} stochastic {Bethe}–{Salpeter} spectra for large systems. \emph{The Journal of Chemical Physics} \textbf{2023}, \emph{158}, 154104\relax
\mciteBstWouldAddEndPuncttrue
\mciteSetBstMidEndSepPunct{\mcitedefaultmidpunct}
{\mcitedefaultendpunct}{\mcitedefaultseppunct}\relax
\EndOfBibitem
\bibitem[Wei\ss{}e \latin{et~al.}(2006)Wei\ss{}e, Wellein, Alvermann, and Fehske]{Weisse2006}
Wei\ss{}e,~A.; Wellein,~G.; Alvermann,~A.; Fehske,~H. The kernel polynomial method. \emph{Rev. Mod. Phys.} \textbf{2006}, \emph{78}, 275--306\relax
\mciteBstWouldAddEndPuncttrue
\mciteSetBstMidEndSepPunct{\mcitedefaultmidpunct}
{\mcitedefaultendpunct}{\mcitedefaultseppunct}\relax
\EndOfBibitem
\bibitem[Neese \latin{et~al.}(2020)Neese, Wennmohs, Becker, and Riplinger]{Neese2020}
Neese,~F.; Wennmohs,~F.; Becker,~U.; Riplinger,~C. The ORCA quantum chemistry program package. \emph{Journal of Chemical Physics} \textbf{2020}, \emph{152}\relax
\mciteBstWouldAddEndPuncttrue
\mciteSetBstMidEndSepPunct{\mcitedefaultmidpunct}
{\mcitedefaultendpunct}{\mcitedefaultseppunct}\relax
\EndOfBibitem
\bibitem[Neese(2022)]{Neese2022}
Neese,~F. Software update: The ORCA program system—Version 5.0. \emph{WIREs Computational Molecular Science} \textbf{2022}, \emph{12}, e1606\relax
\mciteBstWouldAddEndPuncttrue
\mciteSetBstMidEndSepPunct{\mcitedefaultmidpunct}
{\mcitedefaultendpunct}{\mcitedefaultseppunct}\relax
\EndOfBibitem
\bibitem[Troullier and Martins(1991)Troullier, and Martins]{TroullierMartins91}
Troullier,~N.; Martins,~J.~L. Efficient pseudopotentials for plane-wave calculations. \emph{Phys. Rev. B} \textbf{1991}, \emph{43}, 1993--2006\relax
\mciteBstWouldAddEndPuncttrue
\mciteSetBstMidEndSepPunct{\mcitedefaultmidpunct}
{\mcitedefaultendpunct}{\mcitedefaultseppunct}\relax
\EndOfBibitem
\bibitem[Sun \latin{et~al.}(2018)Sun, Berkelbach, Blunt, Booth, Guo, Li, Liu, McClain, Sayfutyarova, Sharma, Wouters, and Chan]{Sun2018}
Sun,~Q.; Berkelbach,~T.~C.; Blunt,~N.~S.; Booth,~G.~H.; Guo,~S.; Li,~Z.; Liu,~J.; McClain,~J.~D.; Sayfutyarova,~E.~R.; Sharma,~S.; Wouters,~S.; Chan,~G. K.~L. PySCF: the Python-based simulations of chemistry framework. \emph{Wiley Interdisciplinary Reviews: Computational Molecular Science} \textbf{2018}, \emph{8}\relax
\mciteBstWouldAddEndPuncttrue
\mciteSetBstMidEndSepPunct{\mcitedefaultmidpunct}
{\mcitedefaultendpunct}{\mcitedefaultseppunct}\relax
\EndOfBibitem
\bibitem[Dunning \latin{et~al.}(2001)Dunning, Peterson, and Wilson]{dunning2001}
Dunning,~T.~H.; Peterson,~K.~A.; Wilson,~A.~K. Gaussian basis sets for use in correlated molecular calculations. {X}. {The} atoms aluminum through argon revisited. \emph{J. Chem. Phys.} \textbf{2001}, \emph{114}, 9244--9253, Covers extensions and optimizations of cc-pVXZ/aug-cc-pVXZ\relax
\mciteBstWouldAddEndPuncttrue
\mciteSetBstMidEndSepPunct{\mcitedefaultmidpunct}
{\mcitedefaultendpunct}{\mcitedefaultseppunct}\relax
\EndOfBibitem
\bibitem[Martyna and Tuckerman(1999)Martyna, and Tuckerman]{MartynaTuckerman1999}
Martyna,~G.~J.; Tuckerman,~M.~E. {A reciprocal space based method for treating long range interactions in ab initio and force-field-based calculations in clusters}. \emph{The Journal of Chemical Physics} \textbf{1999}, \emph{110}, 2810--2821\relax
\mciteBstWouldAddEndPuncttrue
\mciteSetBstMidEndSepPunct{\mcitedefaultmidpunct}
{\mcitedefaultendpunct}{\mcitedefaultseppunct}\relax
\EndOfBibitem
\bibitem[Golub and Businger(1965)Golub, and Businger]{golub1965}
Golub,~G.~H.; Businger,~P. Linear least squares solutions by {Householder} transformations. \emph{Numer. Math.} \textbf{1965}, \emph{7}, 269--276\relax
\mciteBstWouldAddEndPuncttrue
\mciteSetBstMidEndSepPunct{\mcitedefaultmidpunct}
{\mcitedefaultendpunct}{\mcitedefaultseppunct}\relax
\EndOfBibitem
\bibitem[Press \latin{et~al.}(2007)Press, Teukolsky, Vetterling, and Flannery]{press2007}
Press,~W.~H.; Teukolsky,~S.~A.; Vetterling,~W.~T.; Flannery,~B.~P. \emph{Numerical Recipes: The Art of Scientific Computing}, 3rd ed.; Cambridge University Press, 2007; Chapter 2.10, Householder QR decomposition in computational physics\relax
\mciteBstWouldAddEndPuncttrue
\mciteSetBstMidEndSepPunct{\mcitedefaultmidpunct}
{\mcitedefaultendpunct}{\mcitedefaultseppunct}\relax
\EndOfBibitem
\bibitem[Szabo and Ostlund(1996)Szabo, and Ostlund]{szabo1996}
Szabo,~A.; Ostlund,~N.~S. \emph{Modern Quantum Chemistry: Introduction to Advanced Electronic Structure Theory}; Dover, 1996; Chapter 1.4, Gram-Schmidt in quantum chemistry basis sets\relax
\mciteBstWouldAddEndPuncttrue
\mciteSetBstMidEndSepPunct{\mcitedefaultmidpunct}
{\mcitedefaultendpunct}{\mcitedefaultseppunct}\relax
\EndOfBibitem
\bibitem[Daniel \latin{et~al.}(1976)Daniel, Gragg, Kaufman, and Stewart]{daniel1976}
Daniel,~J.~W.; Gragg,~W.~B.; Kaufman,~L.; Stewart,~G.~W. Reorthogonalization and stable algorithms for updating the Gram-Schmidt QR factorization. \emph{Math. Comp.} \textbf{1976}, \emph{30}, 772--795\relax
\mciteBstWouldAddEndPuncttrue
\mciteSetBstMidEndSepPunct{\mcitedefaultmidpunct}
{\mcitedefaultendpunct}{\mcitedefaultseppunct}\relax
\EndOfBibitem
\bibitem[Golub and Reinsch(1970)Golub, and Reinsch]{golub1970}
Golub,~G.~H.; Reinsch,~C. Singular value decomposition and least squares solutions. \emph{Numer. Math.} \textbf{1970}, \emph{14}, 403--420\relax
\mciteBstWouldAddEndPuncttrue
\mciteSetBstMidEndSepPunct{\mcitedefaultmidpunct}
{\mcitedefaultendpunct}{\mcitedefaultseppunct}\relax
\EndOfBibitem
\bibitem[Löwdin(1950)]{lowdin1950}
Löwdin,~P.-O. On the non-orthogonality problem connected with the use of atomic wave functions in the theory of molecules and crystals. \emph{J. Chem. Phys.} \textbf{1950}, \emph{18}, 365--375\relax
\mciteBstWouldAddEndPuncttrue
\mciteSetBstMidEndSepPunct{\mcitedefaultmidpunct}
{\mcitedefaultendpunct}{\mcitedefaultseppunct}\relax
\EndOfBibitem
\bibitem[Bowler and Miyazaki(2012)Bowler, and Miyazaki]{bowler2012}
Bowler,~D.~R.; Miyazaki,~T. Methods in electronic structure calculations. \emph{Rep. Prog. Phys.} \textbf{2012}, \emph{75}, 036503, Covers SVD in DFT applications\relax
\mciteBstWouldAddEndPuncttrue
\mciteSetBstMidEndSepPunct{\mcitedefaultmidpunct}
{\mcitedefaultendpunct}{\mcitedefaultseppunct}\relax
\EndOfBibitem
\bibitem[Bruneval and Gonze(2008)Bruneval, and Gonze]{Bruneval2008}
Bruneval,~F.; Gonze,~X. Accurate $GW$ self-energies in a plane-wave basis using only a few empty states: Towards large systems. \emph{Phys. Rev. B} \textbf{2008}, \emph{78}, 085125\relax
\mciteBstWouldAddEndPuncttrue
\mciteSetBstMidEndSepPunct{\mcitedefaultmidpunct}
{\mcitedefaultendpunct}{\mcitedefaultseppunct}\relax
\EndOfBibitem
\bibitem[Rocca \latin{et~al.}(2008)Rocca, Gebauer, Saad, and Baroni]{Rocca2008}
Rocca,~D.; Gebauer,~R.; Saad,~Y.; Baroni,~S. Turbo charging time-dependent density-functional theory with Lanczos chains. \emph{The Journal of Chemical Physics} \textbf{2008}, \emph{128}, 154105\relax
\mciteBstWouldAddEndPuncttrue
\mciteSetBstMidEndSepPunct{\mcitedefaultmidpunct}
{\mcitedefaultendpunct}{\mcitedefaultseppunct}\relax
\EndOfBibitem
\bibitem[Becke(1988)]{Becke1988}
Becke,~A.~D. A multicenter numerical integration scheme for polyatomic molecules. \emph{The Journal of Chemical Physics} \textbf{1988}, \emph{88}, 2547--2553\relax
\mciteBstWouldAddEndPuncttrue
\mciteSetBstMidEndSepPunct{\mcitedefaultmidpunct}
{\mcitedefaultendpunct}{\mcitedefaultseppunct}\relax
\EndOfBibitem
\bibitem[Becke(1988)]{Becke1988pra}
Becke,~A.~D. Density-functional exchange-energy approximation with correct asymptotic behavior. \emph{Phys. Rev. A} \textbf{1988}, \emph{38}, 3098--3100\relax
\mciteBstWouldAddEndPuncttrue
\mciteSetBstMidEndSepPunct{\mcitedefaultmidpunct}
{\mcitedefaultendpunct}{\mcitedefaultseppunct}\relax
\EndOfBibitem
\bibitem[Dunning(1989)]{Dunning1989}
Dunning,~J.,~Thom~H. Gaussian basis sets for use in correlated molecular calculations. I. The atoms boron through neon and hydrogen. \emph{The Journal of Chemical Physics} \textbf{1989}, \emph{90}, 1007--1023\relax
\mciteBstWouldAddEndPuncttrue
\mciteSetBstMidEndSepPunct{\mcitedefaultmidpunct}
{\mcitedefaultendpunct}{\mcitedefaultseppunct}\relax
\EndOfBibitem
\bibitem[Pritchard \latin{et~al.}(2019)Pritchard, Altarawy, Didier, Gibson, and Windus]{Pritchard2019}
Pritchard,~B.~P.; Altarawy,~D.; Didier,~B.; Gibson,~T.~D.; Windus,~T.~L. New Basis Set Exchange: An Open, Up-to-Date Resource for the Molecular Sciences Community. \emph{Journal of Chemical Information and Modeling} \textbf{2019}, \emph{59}, 4814--4820, PMID: 31600445\relax
\mciteBstWouldAddEndPuncttrue
\mciteSetBstMidEndSepPunct{\mcitedefaultmidpunct}
{\mcitedefaultendpunct}{\mcitedefaultseppunct}\relax
\EndOfBibitem
\end{mcitethebibliography}

\end{document}